\documentclass[runningheads]{llncs}
\usepackage{amsfonts}
\usepackage{graphicx}
\usepackage{hyperref}
\usepackage{xcolor}
\usepackage{subfig}
\usepackage{wrapfig}
\usepackage{floatrow}
\usepackage{fullpage}
\usepackage{framed}
\newtheorem{assumption}[theorem]{Assumption}
\newcommand{\formable}{{feasible}}

\begin{document}
\title{Mobile RAM and Shape Formation by Programmable Particles}
\titlerunning{Programmable Particles}
%
\author{Giuseppe Antonio Di Luna\inst{1} \and
Paola Flocchini \inst{2} \and
Nicola Santoro \inst{3} \and Giovanni Viglietta\inst{4}
\and Yukiko Yamauchi \inst{5}}
\authorrunning{Di Luna et al.}
%
\institute{Sapienza University of Rome, Italy, \email{diluna@diag.uniroma1.it} \and
University of Ottawa, Canada, \email{paola.flocchini@uottawa.ca} \\ \and
  Carleton University, Ottawa, Canada,
\email{santoro@cs.carleton.ca}\\ \and  JAIST, Nomi city, Japan, \email{johnny@jaist.ac.jp} \\ \and Kyushu University, Fukuoka, Japan, \email{yamauchi@inf.kyushu-u.ac.jp}}
\maketitle              

\begin{abstract}
We investigate computational issues in the  distributed   model  \emph{Amoebots}  
of programmable matter. In this model, the computational entities, called particles,  are anonymous 
finite-state machines that operate and  move on a hexagonal tessellation of the plane.

In this paper we show how  a constant number of such weak particles can 
simulate  a powerful  Turing-complete entity that is able to move on the plane while computing. 

We then show an application of our tool to the classical  Shape-Formation problem,
providing a new and much more general distributed solution protocol.
Indeed, the existing algorithms  would allow to form only shapes made of arrangements of segments and triangles.
Our algorithm allows the particles to form more abstract and general 
connected shapes,  including  circles and spirals, as well as  fractal objects of non-integer dimension, such as the Sierpinski triangle or the Koch snowflake.

In lieu of the existing impossibility results based on
 the symmetry of the initial
configuration of the particles, our result provides a   
 complete characterization of the connected shapes that can be formed by an initially simply connected set of particles. Furthermore, in the case of non-connected shapes, we give almost-matching necessary and sufficient conditions for their formability.
\end{abstract}

\section{Introduction}\label{s:intro}

Several  parallel  and distributed computing  models have been devoted to  formalizing  
computations within the
interdisciplinary field of Programmable Matter  (PM): see \cite{xmichail2,xnewref,xPa14,xRo06,xScW15,xWaWA04}.
The PM field envisions a myriad of very small (micro/nano-sized) entities  that are 
nevertheless able to move and coordinate themselves with the final purpose of solving a specific task \cite{xToM91}.
Practical research is designing, engineering, and developing prototypes that will lead to  future hardware platforms for PM. Examples are the M-blocks: cubes that are able to rearrange themselves by rotations \cite{6696971}, and the Kilobots: small robots that move by vibrations   \cite{6224638}.
At the same time, the algorithmic community is formalizing abstract and general models that capture the peculiarities behind these hardware platforms,  enabling the development of provably correct algorithms 
and the feasibility analysis of problems.

 In this paper, we consider the popular geometric {\em Amoebot} model, 
 introduced in \cite{xDerGSB+15}.
In the Amoebot model, a set of computationally limited identical  entities, 
called {\em  particles}, 
operate and move on a hexagonal tessellation of the plane (i.e., a triangular grid). Each particle 
has constant-size memory (i.e., constant with respect to the total number of particles), is anonymous (i.e., it has no ID), is able to communicate only with its direct neighbors in the grid, and it moves by repeating an 
{\em expansion action} (in which the particle expands to occupy two neighboring nodes of the grid) and a {\em contraction action} (in which an expanded particle contracts to a single node of the grid). 
Research using this model is being carried out  within  the parallel, distributed, and molecular computing fields
 (e.g., see \cite{ArCDRR18,xCaDRR16,xDaGPRSS18,xDaGPR+17,xDerGMR+15,xspaa,xDerGMR+17,xDerGSB+15,previous,xlinerec}; for a recent survey, see \cite{DaHRS19}). 
The main goal of these research  efforts 
is to gain an understanding of the nature and limits of this distributed computational universe.

In this paper, we  move one step forward in this quest  by providing a construction that simulates a 
{\em moving  Random-Access Machine} (mRAM) using four particles. Such a construction transforms a set of these weak particles into a powerful  Turing-complete entity that is able to move on the grid while computing. We prove the usefulness of our construction by applying it to the well-studied  
{\em Shape-Formation} (or Pattern-Formation) problem.

 In the Shape-Formation problem, the particles, initially arranged in an arbitrary 
 connected shape, 
 have to form a given target shape.   More precisely, each particle starts with a representation of the target shape in its memory, and coordinates with the other particles to form a suitably scaled-up copy of the shape that includes all particles in the system (this could mean that some particles have to be in the expanded state in the final shape). Usually, the total number of particles $n$ is unknown: as a matter of fact, $n$ cannot be stored in the constant-size memory of a single particle.
 
In a series of works \cite{xDerGMR+15,xspaa,DerGSB+15,previous}, 
 increasingly refined and complex techniques and algorithms have been designed, each enlarging 
  the class of shapes that can be formed starting from a simply connected configuration. 
To date,  however, this class includes only target shapes defined as an arrangement of segments and triangles. 

\begin{figure}[h!]
\begin{center}
\includegraphics[height=8cm]{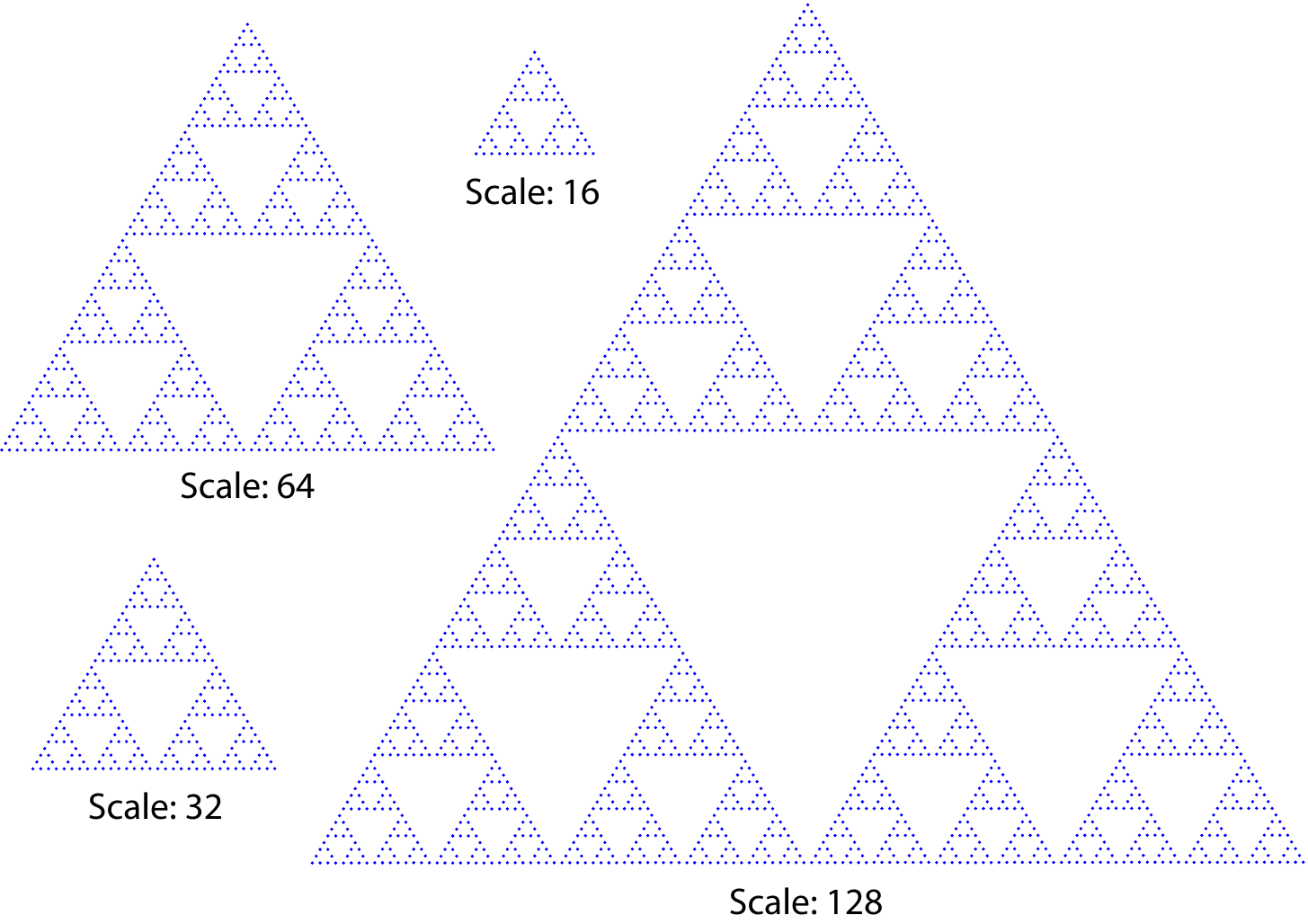}
\end{center}
\caption{A Sierpinski triangle at various scales, approximated by a system of particles}
\label{f:siepinski}
\end{figure}

Thanks to our mRAM simulation, we are able to provide a universal shape-formation algorithm where the target shape is any subset of the grid whose scaled-up versions can be generated by an algorithm. 
In other words,  we do not enforce a geometric relationship between scaled-up versions of the same shape: the final shape could be a curve, such as a circle or a spiral, or a more complex fractal object
 of non-integer dimension, such as the Sierpinski triangle or the Koch snowflake (see Fig.~\ref{f:siepinski}).

Therefore, our second contribution is a general and universal solution for the Shape-Formation problem: starting from a simply connected shape (i.e., a shape without holes),  our algorithm
allows the particles to form any \formable\ {\em connected} shape for which a ``drawing algorithm'' exists (i.e., the shape is Turing-computable). Such a drawing algorithm is a RAM algorithm that takes as input the number of particles $n$, and outputs (a representation of) the shape that has to be formed by the $n$ particles. Note that, since the number of instructions of such a drawing algorithm is obviously constant with respect to $n$,\footnote{By contrast, the \emph{running time} of the drawing algorithm may be any function of $n$.} the algorithm itself can fit in the constant-size memory of a single particle, regardless of the number of particles in the system. (By comparison, previous works
  assumed that each particle  has in memory a representation of the target shape expressed as a constant number of segments and triangles  \cite{xDerGMR+15,xspaa,DerGSB+15,previous}.)
  
  With our technique, we can form almost all ``feasible''  {\em non-connected} target shapes, and are excluded from our result only very sparse pathological  shapes. 
With regards to the  {\em feasibility} (or not) of   a shape,
it is known that, depending  on  the symmetry of the initial configuration,
some shapes  are not formable 
 regardless of the amount of memory \cite{previous}.
The negative result of \cite{previous} and the positive result of our paper give an almost 
complete characterization of shapes that are formable starting from a simply connected configuration. Interestingly, in the case of \emph{connected} shapes, the characterization is complete. 

Our distributed algorithm is deterministic, and it works even if the schedule of activations is fair  but adversarial. More precisely, in each stage,   upon activation, a particle exchanges messages with its neighbors, executes some computation, and  possibly  moves;  there is however no   restriction on the number of particles that can be concurrently active in the same stage. Moreover, our algorithm works also when the particles do not have {\em chirality} (i.e., there is no common notion of a clockwise direction on the plane among the particles). 

 
%

\paragraph{Paper outline.} In Section \ref{s:definition} we present the formal model and the Shape-Formation problem. In Section \ref{s:RAM} we discuss how to simulate a mobile RAM using four particles.
In Section \ref{s:old} we summarize the initial phases of the shape-formation algorithm used in \cite{previous}; our algorithm leverages the same initial phases that are used to arrange particles in a specific configuration.  Finally, in Section \ref{s:new}, we present our shape-formation algorithm. 
\paragraph{Related work.}
The Shape-Formation problem, in the context of PM, has been first defined in \cite{xDerGMR+15}, where the goal was to form simple shapes such as a hexagon or a triangle. A subsequent work \cite{xspaa} has shown how to form target shapes defined as arrangements of triangles. However, both works where assuming {\em chirality} and a sequential activation scheduler (at most one particle active at each time). Finally, \cite{previous} completely characterized the target shapes, defined as arrangement of triangles and segments, that can be formed from a simply connected initial configuration. This last result holds in a system with no chirality and under an adversarial scheduler.

A key tool of the algorithm in \cite{previous} is the simulation of a Turing machine. Such a tool is similar in spirit to our mRAM simulation; however, it has a shortcoming: it simulates a Turing machine using a line of connected particles (on this line, a special leader particle acts as the Turing machine's head), and therefore the size of this Turing machine's tape is limited by the number of particles on the line. Our new construction does not have this shortcoming, because it needs only a constant number of particles to simulate any Turing machine.

In other models, some works \cite{dilunaline,michail} have investigated the possibility of using weak entities to simulate a more powerful one by implementing a Turing machine, but their models differ substantially from ours, and their methods cannot be carried over.




\section{The Amoebot model and the Shape-Formation problem}\label{s:definition}
\subsection{Particles}

The Amoebot model we use in this paper is the one described in~\cite{previous}, which we summarize here. In this model, a \emph{particle} is a computational entity that lives in an infinite regular triangular grid $G$ embedded in the Euclidean plane $\mathbb{R}^2$ (observe that a regular triangular grid corresponds to a hexagonal tessellation of the plane). A particle may occupy either one vertex of $G$ or two adjacent vertices: in the first case, the particle is said to be \emph{contracted}; otherwise, it is \emph{expanded}. When it is expanded, one of the vertices it occupies is called its \emph{head}, and the other vertex is its \emph{tail}. A particle may move through $G$ by repeatedly expanding toward a neighboring vertex of $G$ and contracting into its head. (The traditional Amoebot model also includes a special type of coordinated move called ``handover'', but we will not need it in this paper.

No vertex of $G$ can ever be occupied by more than one particle at a time. So, a contracted particle cannot expand into a vertex that is already occupied by another particle. Also, if two or more particles attempt to expand toward the same (unoccupied) vertex at the same time, only one of them succeeds, chosen arbitrarily by an adversary.

At each \emph{stage}, some particles in the system are \emph{active}, and they perform a \emph{look-compute-move cycle}, and the other particles are \emph{inactive}. An adversarial \emph{  scheduler} arbitrarily and unpredictably decides which particles are active at each stage. The only restriction on the scheduler is that it cannot keep a particle inactive forever, but it must activate every particle infinitely often.

When a particle is activated for a certain stage, it ``looks'' at the vertices of $G$ adjacent to its head, discovering if they are currently unoccupied, or if they are head or tail vertices of some particle. All particles are \emph{anonymous} (i.e., they are indistinguishable). Each active particle may then decide to expand, contract, or stay still for that stage. When the next stage starts, a new set of active particles is selected, which observe their surroundings and move, and so on.

Each particle has an \emph{internal state} that it can modify every time it is activated. The internal state of any particle must be picked from a finite set: particles have an amount of ``memory'' that is constant with respect to the size of the system, $n$.

Two particles can also \emph{communicate} by sending \emph{messages} to one another. Each message is taken from a finite set of predefined messages. An active particle can send a message to another particle provided that their heads are adjacent vertices of $G$. A particle reads the incoming messages from all its neighbors as soon as it is activated.

Each particle labels its six incident edges with \emph{port numbers}, going from $0$ to $5$. Each particle uses a consistent numbering that is invariant under translation on $G$. However, different particles may disagree on which of the edges has port number $0$ and whether the numbering should follow the clockwise or counterclockwise order: this is called the particles' \emph{handedness}. So, the handedness of a particle does not change as the particle moves, but different particles may have different handedness.

At each stage, each active particle looks at its surroundings to see which neighboring vertices are occupied, and it reads the incoming messages. Based on these and on its internal state, the particle executes a \emph{deterministic algorithm} that computes a new internal state, the messages to be sent to the neighbors, and whether the particle should expand to some adjacent vertex, contract, or stay still.

\subsection{General definition of shape}
A set $S$ of nodes of $G$ is \emph{formable by $n$ particles} if there exists a configuration of exactly $n$ particles (contracted or expanded) that collectively occupy exactly the nodes in $S$.

A \emph{shape} is a function mapping a positive integer $n$ to a set $S_n$ of nodes of $G$ that is formable by $n$ particles: this set $S_n$ is called the $n$th \emph{level} of the shape. If such a function is Turing-computable, then the shape is said to be \emph{computable}. If every level of a shape is a connected set, the shape is said to be \emph{connected}.

A shape is \emph{formable} (under condition $\mathcal C$) if there exists a distributed algorithm that, for every $n$, makes any system of $n$ particles (whose initial configuration satisfies condition $\mathcal C$) eventually form a copy of the $n$th level of the shape, possibly translated, rotated by a multiple of $60^\circ$, and reflected. The algorithm should succeed regardless of the port numbers of each particle and the choices of the adversarial scheduler, and it should guarantee that the system remains still after forming the shape.

\subsection{Unbreakable symmetries}
We borrow from~\cite{previous} the concept of ``unbreakable symmetry''. A configuration of particles is said to be \emph{unbreakably $k$-symmetric}, for some integer $k>1$, if it has a center of $k$-fold rotational symmetry that does not coincide with any node of $G$. Observe that there exist unbreakably $k$-symmetric configurations only for $k=2$ and $k=3$.

We can extend this notion to a level of a shape in a natural way: the $n$th level of a shape is unbreakably symmetric if it is formable by an unbreakably $k$-symmetric configuration of $n$ particles.
As noted in~\cite{previous}, if the system initially is in an unbreakably $k$-symmetric configuration, the port labels of symmetric particles happen to be symmetric, and the scheduler always decides to activate symmetric particles simultaneously, then the system never ceases to be in an unbreakably $k$-symmetric configuration. Intuitively, the absence of a central node (and therefore of a central particle) and the fact that the configuration is initially symmetric makes it impossible to break the symmetry. Hence, we have a \emph{necessary} condition $\mathcal C$ for the formability of a shape:
\begin{proposition}\label{p:necessary}
A shape is formable only under the condition that, if its $n$th level is not unbreakably $k$-symmetric, then the initial configuration of a system of $n$ particles seeking to form the shape should not be unbreakably $k$-symmetric, either.
\end{proposition}

In Section~\ref{s:new} we will give a strong \emph{sufficient} condition for the formability of a shape, which, together with Proposition~\ref{p:necessary}, almost characterizes the set of computable shapes that can be formed from a simply connected initial configuration of particles (in the case of \emph{connected} computable shapes, it yields a full characterization).

\section{Simulating Random-Access Machines}\label{s:RAM}

\subsection{Random-Access Machines: definition}
A \emph{Random-Access Machine} (RAM) is a model of computation consisting of a finite set of \emph{registers}, each of which can store a non-negative integer, and a \emph{program} consisting of a finite ordered sequence of \emph{instructions}. Each instruction is of one of two types:
\begin{itemize}
\item ${\rm Inc}(r)$: increment by $1$ the value stored in the register $r$. Then, proceed to the next instruction of the program.
\item ${\rm TestDec}(r, i)$: if the register $r$ is holding the value $0$, jump to the $i$th instruction of the program. Otherwise, decrement $r$ by 1 and proceed to the next instruction.
\end{itemize}
The registers initially contain the \emph{input} of the RAM, and then the program is executed starting from the first instruction. We can reserve a register $r_t$ to represent a ``termination flag'', whose value is initially $0$. When $r_t$ is incremented, the value of the other registers is taken as the RAM's \emph{output}. Hence, a RAM is a device that can compute integer functions.

\subsection{Simulating Turing Machines by RAMs}
In~\cite[Chapter~11]{Minsky}, Minsky shows how RAMs can simulate Turing Machines. Specifically, there is a (small) constant $c$ such that, given any Turing Machine $T$ that computes a function $f_T$, there exists a RAM $Q_T$ with exactly $c$ registers that computes $f_T$.

Then, in~\cite[Chapter~14]{Minsky}, he shows that any RAM $R$ can be simulated by a RAM $Q'_R$ having only two registers. This is done by encoding the set of values stored in $R$'s registers as a single integer, which is stored in the first register of $Q'_R$ (the second register of $Q'_R$ is only used for intermediate computations). The code is based on G\"odel numbers: for instance, the sequence $(a_1, a_2, a_3, a_4, a_5)$ is encoded as the single integer $2^{a_1}\cdot 3^{a_2}\cdot 5^{a_3}\cdot 7^{a_4}\cdot 11^{a_5}$. In general, the $i$th integer in the sequence becomes the exponent of the $i$th prime factor of the code. By the unique-factorization theorem, this encoding is injective and therefore non-ambiguous.

The program of $Q'_R$ can be constructed by locally replacing each instruction of $R$ with a small program that simulates it. For instance, it is possible to increment the $i$th register of $R$ by multiplying the first register of $Q'_R$ by the $i$th prime number, which in turn can be done in $Q'_R$ by using the two standard instructions and the auxiliary register. Testing if the $i$th register of $R$ is non-$0$ amounts to testing if the first register of $Q'_R$ holds a multiple of the $i$th prime number, etc. In this paradigm, we can stipulate that $Q'_R$ terminates when its second register holds a $0$ and the value of its first register is a multiple of the prime number corresponding to the termination flag of $R$.

As an immediate consequence of the above, we have that the RAMs with only two registers can simulate all Turing Machines, and can therefore compute any computable function.

\subsection{Simulating RAMs by particles}

\begin{figure}[h!]
\begin{center}
\includegraphics[scale=0.9]{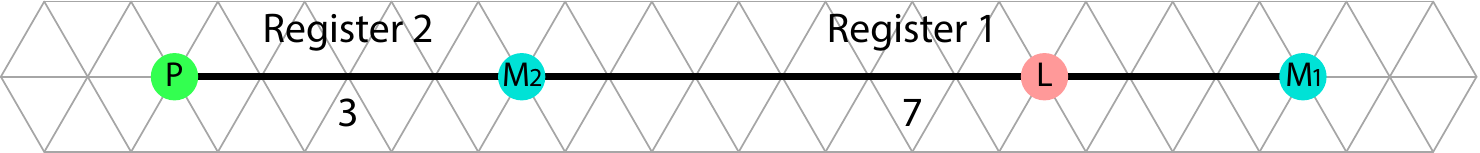}
\end{center}
\caption{A RAM simulator with the first register holding the value $7$ and the second register holding the value $3$.}
\label{f:ram01}
\end{figure}

Our goal in this section is to simulate a RAM with two registers by a set of four particles. The layout of our simulator is shown in Fig.~\ref{f:ram01}: the four particles always remain collinear throughout the simulation, and they always maintain their order along the line: first the \emph{pivot} $P$, then the \emph{marker of the second register} $M_2$, then the \emph{leader} $L$, and finally the \emph{marker or the first register} $M_1$. The number of empty locations between $P$ and $M_2$ (i.e., their distance minus $1$) represents the value stored in the second register of the RAM, and the number of empty locations between $M_2$ and $M_1$ (i.e., their distance minus $2$) represents the value stored in the first register of the RAM.

At all times during the simulation of the RAM's program, $L$ will remember the index of the instruction that is currently being simulated: since the program of a RAM is finite, $L$ only needs a finite amount of memory to do so. Now we have to show that such a system can simulate every possible instruction of the RAM. This is done by letting $L$ move between $M_2$ and $M_1$: we assume that $L$ knows in which direction it has to move to find each of these two particles, and for convenience we call these directions ``left'' and ``right'', respectively, to match Fig.~\ref{f:ram01}. When $L$ reaches the relevant particle, it communicates with it and causes it to move according to the current instruction of the program. $P$, on the other hand, always remains still. Specifically, here is how each instruction is simulated:
\begin{itemize}
\item ${\rm Inc}({\rm Register\ }1)$: $L$ moves toward $M_1$ until it finds it. Then it gives $M_1$ the order to move one step to the right, and waits until $M_1$ has moved.
\item ${\rm TestDec}({\rm Register\ }1, i)$: if $L$ neighbors both $M_2$ and $M_1$, it does nothing (and updates the index of the current instruction to $i$). Otherwise, it reaches $M_1$ and orders it to move one step to the left. Then $L$ itself moves one step to the left and waits until $M_1$ has moved.
\item ${\rm Inc}({\rm Register\ }2)$: $L$ reaches $M_1$ and orders it to move one step to the right. When $M_1$ has moved, $L$ reaches $M_2$ and orders it to move one step to the right. Then $L$ itself moves one step to the right and waits for $M_2$.
\item ${\rm TestDec}({\rm Register\ }2, i)$: $L$ reaches $M_2$ and asks it if it has a neighbor on the opposite side (i.e., $P$). If $M_2$ answers affirmatively, $L$ does nothing (and updates the index of the current instruction to $i$). Otherwise, it orders $M_2$ to move one step to the left and waits until it has moved; then it reaches $M_1$ and orders it to move one step to the left; finally, $L$ itself moves one step to the left and waits until $M_1$ has moved.
\end{itemize}

\subsection{Adding control registers for mobility}
In our Shape-Formation algorithm, we will need the RAM simulator outlined above to be able to programmatically move and occasionally perform actions depending on the shape to be formed. In this section we outline the mechanism that our RAM we will use to move, and the specific actions needed for the Shape-Formation algorithm that will be explained in next sections. 

Recall that, in order to simulate a generic Turing Machine $T$, the RAM $Q_T$ needs only a constant number $c$ of registers. We can augment $Q_T$ by adding a fixed number $c'$ of ``flag registers'', which are set to $1$ when the system has to perform certain special operations.

Specifically, assume that $T$ is an algorithmic description of a shape, whose output is a sequence of \emph{plotting operations} of the form ``move forward'', ``move right'', ``draw a point'', etc. Our mobile RAM (mRAM) $U_T$ simulates $T$ exactly like $Q_T$, with one exception: whenever $T$ outputs a plotting operation, $U_T$ sets and then immediately resets the flag register corresponding to that type of operation (the reason for this will be explained shortly).

Now, we let the mRAM $Q'_{U_T}$ simulate $U_T$ using only two registers. In $Q'_{U_T}$, the state of each of the $c'$ flag registers of $U_T$ can be checked by verifying, at the end of a simulated instruction, if the value stored in the first register is a multiple of the prime number associated with the flag register.

As our four-particle system simulates $Q'_{U_T}$, it can also test the $c'$ flag registers of $U_T$ at the end of every simulated instruction. Indeed, the leader particle $L$ can test if the value stored in the first register is a multiple of a given prime $p$. It can do so by first moving next to $M_1$, and then counting modulo $p$ the number of steps it takes to move all the way to $M_2$ (counting modulo $p$ requires only $p$ states). Since $c+c'$ is a finite constant, $L$ only ever needs to test a constant set of primes to determine the states of all the flag registers, which in turn takes a finite amount of memory and time.

When $L$ determines that one of the $c'$ flags is set, it executes the corresponding plotting operation of $T$. This translates into a ``movement operation'', which moves the whole system in some direction or orders a specific particle to remain still forever, marking a point of the shape. The exact nature of these movement operations and the details of their implementation will be described in next sections.

\section{Basic shape-formation algorithm}\label{s:old}
The first part of our shape-formation algorithm is taken from~\cite{previous}, while the second part is entirely different, and will be described in Section~\ref{s:new}. Next we will outline the relevant parts of the ``basic algorithm'' of~\cite{previous}.

One assumption of the basic algorithm, which is also an assumption of our new algorithm, is that the initial configuration of the particles is \emph{simply connected}, i.e., it is connected and it has no ``empty holes''. Another assumption is that, if the initial configuration is unbreakably $k$-symmetric, then also the shape to be formed is unbreakably $k$-symmetric. The latter assumption is necessary, due to Proposition~\ref{p:necessary}.

As we pointed out in Section \ref{s:definition}, the basic algorithm only deals with shapes that are made of full triangles and segments, but this assumption is not used in the parts of the basic algorithm that we are going to borrow here. The relevant ``phases'' of the basic algorithm are as follows:
\begin{itemize}
\item \emph{Handedness agreement}: all the particles assume the same handedness (by simply setting an internal flag whose meaning is either ``my original handedness is correct'' or ``my original handedness is incorrect'').
\item \emph{Leader election}: the particles attempt to elect a single leader. If the initial configuration is unbreakably $k$-symmetric, they may fail to do so, and elect exactly $k$ leaders instead.
\item \emph{Straightening}: the particles arrange themselves in $k$ straight line segments, each with a leader located at an endpoint. If $k>1$, all the leaders are pairwise adjacent (recall that the only possibilities are $k=2$ and $k=3$), and the configuration is unbreakably $k$-symmetric (hence the $k$ line segments have the same length and form angles of $2\pi/k$).
\end{itemize}

In the next section we are going to show how to proceed from here: we have a configuration consisting of $k$ straight lines and $k$ leaders, where all particles have the same handedness, and we want to reach an arbitrary configuration (i.e., a level of a computable shape) that is unbreakably $k$-symmetric if $k>1$. 

\section{Forming Turing-computable shapes}\label{s:new}
\subsection{Starting configuration}
\begin{figure}[h!]
\begin{center}
\includegraphics[scale=0.6]{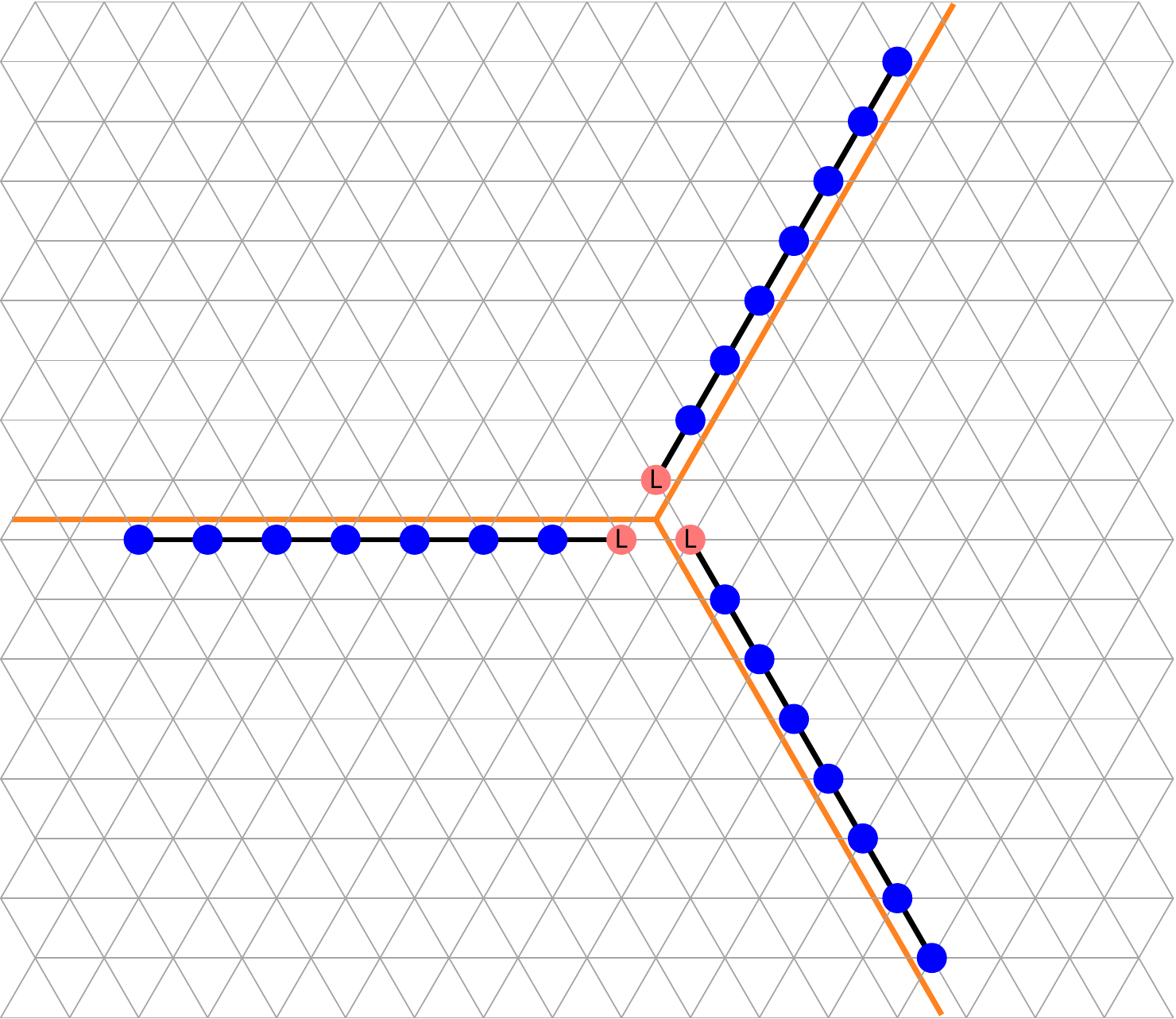}
\end{center}
\caption{The final configuration of the basic algorithm, which is the starting point $C_0$ of our new algorithm. Each leader is assigned a trail of followers, and will guide them in the formation of the part of shape that falls into its principal sector.}
\label{f:form01}
\end{figure}

Recall from Section~\ref{s:old} that, starting from a simply connected configuration, all the particles in the system can agree on a common handedness and rearrange themselves to form a configuration $C_0$ consisting of $k$ equal line segments (with $k\in\{1, 2, 3\}$) each containing a leader particle. Moreover, if $k>1$, then $C_0$ is unbreakably $k$-symmetric, and we may assume that also the shape's level to be formed is unbreakably $k$-symmetric. In this case, each leader is assigned an equal portion of the shape, corresponding to a sector of plane spanned by an angle of $2\pi/k$. These $k$ sectors are called \emph{principal sectors} of the plane, and the $k$ rays separating them are the \emph{principal rays}: Fig.~\ref{f:form01} shows an example for $k=3$.

As a preliminary move, each leader will reach the far end of the line segment on which it is located. This is done by repeatedly ``transferring the leadership'' to a neighboring particle. In turn, this amounts to sending a special message to that neighbor, whose meaning is ``you are now the leader, and I am a regular particle''. So, no particle will actually move in this phase.

When a leader has reached the far end of the segment, it starts the next phase, which consists in building and initializing a mRAM simulator that will eventually form the portion of shape that falls into that principal sector. From this time on, the $k$ leaders will act independently of each other, never meeting again and never interacting.

\subsection{From a shape-generation algorithm to a tracing mRAM}

As stated in Section~\ref{s:definition}, we assume that the shape to be formed is computable, i.e., there exists an algorithm $A$ that generates its $n$th level $S_n$ given the number $n$ as input. By this we mean that $A$ outputs the coordinates of all the points of $S_n$, in an arbitrary order and in some arbitrary coordinate system, and then terminates.

Recall that, by definition of shape, there exists a configuration $C_f$ of $n$ particles, expanded or contracted, which collectively form $S_n$. Additionally, due to Proposition~\ref{p:necessary}, if $k>1$, then $S_n$ must be unbreakably $k$-symmetric, and therefore we may take $C_f$ to be an unbreakably $k$-symmetric configuration of particles. As such, $C_f$ can be translated and rearranged so that its center coincides with the center of $C_0$. Also, $C_f$ can be decomposed into $k$ equal subsets, called \emph{principal subsets}, each of which entirely lies in a principal sector of $C_0$, with one exception: some expanded particles of $C_f$ may be crossing a principal ray, and therefore have the head in a principal sector and the tail in another. These will be called \emph{trespassing particles}. Now, from $A$ we can easily produce a modified algorithm $A'$ that only outputs the points that are occupied by one of the principal subsets of $C_f$, say, $C'_f$.

Furthermore, given $A'$, we can construct another algorithm $A''$ that ``traces'' $C'_f$. That is, $A''$ output the points of $C'_f$ in a specific order: it starts generating the points of $C'_f$ that lie next to a principal ray, from the closest to the principal ray's endpoint to the farthest, and then proceeds to the next ray parallel to the principal ray, and so on (each of these rays is called a \emph{scanline}). This is done by taking every point $p$ on every scanline, in order, and executing $A'$. If $A'$ generates $p$, then $p$ is generated by $A''$. If $A'$ generates points that appear on the same scanline but after $p$, then $A''$ proceeds along the scanline by one step and executes $A'$ again, etc. Otherwise, $A''$ moves to the first point of the next scanline, etc. As soon as $A'$ generates only points that lie on already-visited scanlines, $A''$ terminates. Hence, $A''$ is a procedure that terminates in a finite amount of time.

\begin{figure}[h!]
\center
\includegraphics[scale=0.8]{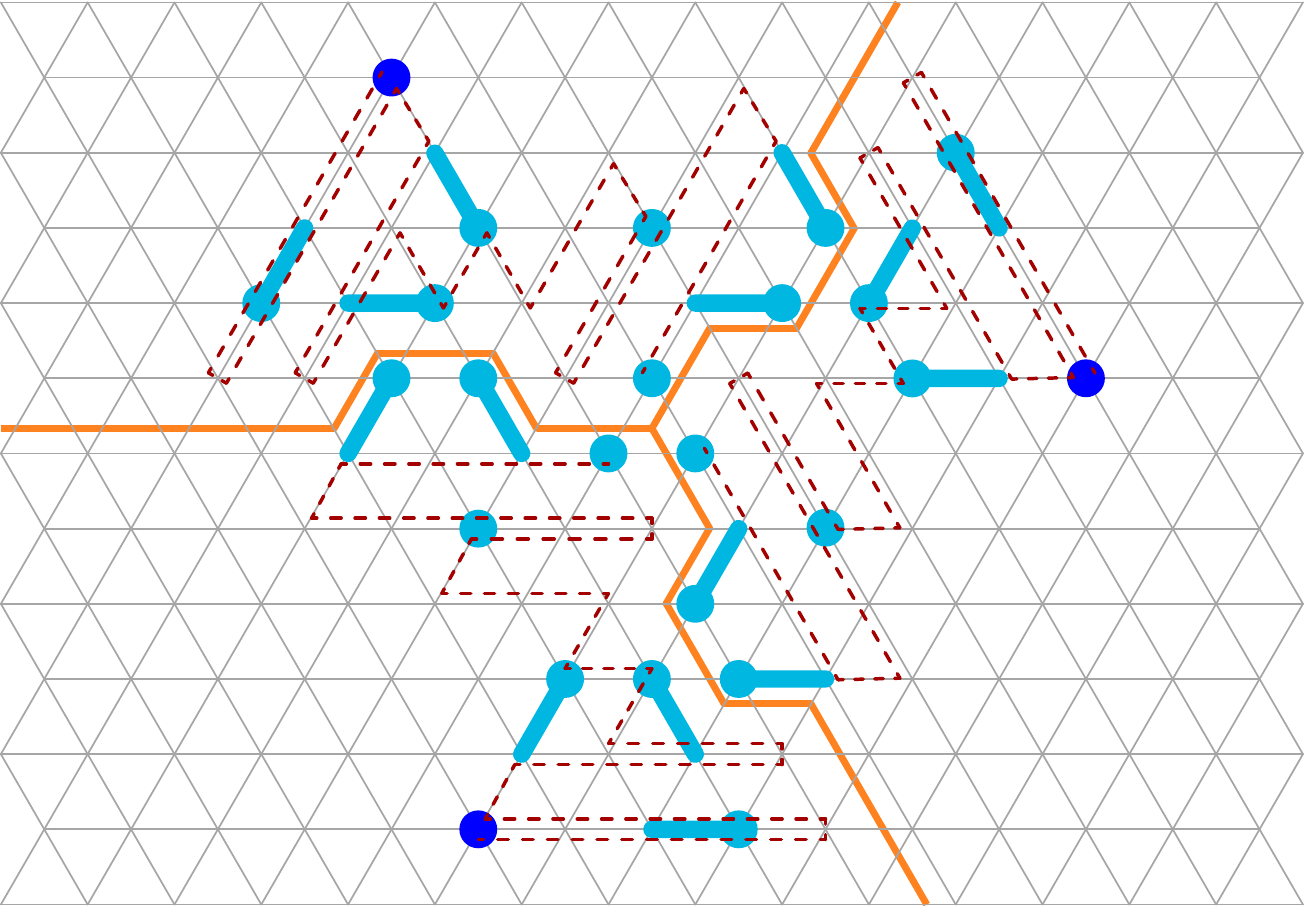}

\caption{A configuration forming the $n$th level of a shape, subdivided into its three principal subsets, each having two trespassing particles. The dashed lines represent the order in which the principal subsets are traced by the algorithm $A''$. The dark particles represent the last dots of their respective principal subsets.}
\label{f:form02}
\end{figure} 
As $A''$ executes $A'$, it can also detect if each location $p$ it generates should be occupied by a single particle or by an expanded particle (indeed, this information is part of the description of $C'_f$). If it is an expanded particle, and the other location $p'$ occupied by the same particle is on the next scanline, $A''$ does not generate $p$: it will generate only $p'$ as soon as it scans it. Along with $p'$, $A''$ will also generate a ``direction of expansion'', which goes from $p'$ to $p$. This is called the \emph{delayed-deployment rule}, and its purpose will be explained later.

Another simple improvement we can make on $A''$ is that it not only generates the locations of the particles of $C'_f$, but it also generates the ``movements'' it makes through the plane to generate them: recall that we introduced these \emph{movement operations} at the end of Section~\ref{s:RAM}. Specifically, the possible movement operations are: (1) advance by one step along the current scanline; (2) move to the next scanline; (3) go back by one step along the current scanline; (4) deploy a contracted particle; (5) deploy a particle that will expand in direction $d$; (6) terminate.

The third type of operation is repeatedly used right after moving to the next scanline, in order to reach its endpoint $e$ and then resume scanning. The only caveat is that, if $e$ should be occupied by a trespassing particle belonging to $C_f\setminus C'_f$, it is skipped by $A''$, and the scan will not be resumed from $e$ but from the location right before it. This type of behavior is called \emph{trespasser-avoidance rule}, and is illustrated in Fig.~\ref{f:form02}.

Since we have an algorithm $A''$, there exists a Turing Machine $T$ that computes the same function: it takes an integer $n$ as input, and it outputs a sequence of movement operations that trace (a sector of) $S_n$. From Section~\ref{s:RAM} we know that there is a  RAM $Q_T$ with $c$ registers that produces the same output when the number $n$ is initially stored in its first register. We can then construct the mobile mRAM $U_T$, with $c'=6$ flag registers, each of which corresponds to one of the movement operations above, and is set and reset whenever the corresponding operation has to be performed.

Finally, we can construct the mRAM $Q'_{U_T}$, which simulates $U_T$ using only two registers, provided that the value $2^n$ is initially stored in its first register. This mRAM will be simulated by each leader in the system and its trail of particles. In the rest of this section we will show how the RAM simulator of Section~\ref{s:RAM} can be expanded to implement the movement operations above and therefore yield a shape-formation algorithm.

\subsection{Initializing the machine}
Let us focus on a single leader particle $L$ and its trail of $n/k$ particles. Recall that the leader $L$ is now located at the endpoint of its trail of particles that is farthest from the center of $C_0$. We want the last four particles on the trail, including $L$, to start executing the simulator of $Q'_{U_T}$ as described in Section~\ref{s:RAM}. First, however, it is necessary to initialize the simulator by storing the value $2^n$ in its first register, which amounts to placing particle $M_1$ at a distance of $2^n+2$ steps away from particle $M_2$ (refer to Fig.~\ref{f:ram01}).

At first $L$ switches states with its only neighbor, which now becomes the new leader $L$. The two neighbors of $L$ become $M_1$ and $M_2$, and then $M_2$ sends a message to its other neighbor, which becomes the pivot $P$.

Then, $L$ has to ``push'' $M_1$ exactly $2^n$ times in order to create the input to the simulator. By ``push'' we mean that $L$ reaches $M_1$, orders it to move forward by one step, and waits until $M_1$ has expanded and then contracted again.

In order to repeat this action the right number of times, the entire trail counts in binary from $0$ to $2^n-1$: this is done by letting each particle hold $k$ binary digits, initially all set to $0$. After each push, $M_1$ adds $1$ to the binary number formed by its $k$ digits. $L$ reaches $M_1$ and takes the carry bit of the addition from $M_1$. If the carry bit is $1$, $L$ adds $1$ to its own $k$ digits. If the carry bit of this addition is $1$, $L$ moves all the way to $M_2$ and orders it to increment its $k$ bits by $1$, etc. As soon as a carry bit is $0$, a message is sent back to $L$ and forwarded by the trail's particles. When the message reaches $L$, it proceeds with the next push operation, etc.

When the last particle in the trail gets a carry bit of $1$, it knows that $M_1$ has been pushed exactly $2^n$ times, and it forwards this information to $L$, which therefore stops pushing and proceeds with the simulation.

\subsection{Tracing the shape}
Let the last particle in the trail be called the \emph{rear particle}. The leader will coordinate the simulation of the mRAM that traces the shape pretending that the ``pen'' is held by the rear particle. That is, the simulator will go on with the computation until a movement operation is reached. If such operation is of type~4 or~5 (i.e., the deployment of a particle), then the rear particle is ``dropped'', which corresponds to drawing a point of the shape.

As explained in Section~\ref{s:RAM}, $L$ knows if a movement operation has to be performed by checking, at the end of every simulated instruction of the mRAM, if the value of the first register is a multiple of some prime number and the value of the second register is $0$. This implies, in particular, that when a movement operation is executed, the particles $P$ and $M_2$ are next to each other.
 
In the following we will explain in greater detail how the system behaves to implement the six possible movement operations.
\begin{enumerate}
\item Advancing by one step means that the four-particle mRAM simulator, as well as the whole trail of particles up to the rear particle, has to move by one step along the current scanline. First, $L$ moves next to $M_1$ and orders it to advance by one step, waiting until it has moved. Then it goes next to $M_2$ and gives it the same order. $L$ then moves away from $M_2$ by one step and waits for it. In the meantime, $M_2$ gives the same order to $P$ and then advances toward $L$. $P$ does the same as $M_2$, and so on. When the rear particle is reached by the message, it advances and sends an ``all done'' message to its predecessor, which forwards it to $L$. Then, $L$ knows that it may proceed with the simulation.
\item Moving to the next scanline means that the entire trail and four-particle mRAM simulator must move ``sideways'' by one step, in order to relocate themselves on the next scanline $s$. The structure of this operation is similar to the previous one. $L$ moves to $M_1$ and orders it to move sideways onto $s$. There can be no misunderstanding on the direction of movement, since we are assuming that all particles have already reached an agreement on a common handedness. $L$ moves waits until $M_1$ has moved, and then goes to $M_2$ and gives it the same order. $M_2$ forwards the order to $P$, which forwards it to the next particles, etc. When the message reaches the rear particle, it forwards an ``all done'' message to $L$, and then moves onto $s$. When any particle receives the ``all done'' message, it waits until its predecessor has moved onto $s$, then it forwards it to its successor and moves onto $s$. When $L$ has finally moved onto $s$, the procedure is over.
\item To go back along the current scanline, the operations of item~$1$ are executed in reverse order. That is, $L$ moves to $M_2$ and forwards a message to the rear particle, which then moves backwards, and all other particles up to $L$ follow one by one. Then $L$ goes to $M_1$ and orders it to move backwards.
\item Deploying a contracted particle means that the rear particle has to stop where it is and remain there forever. This is simple to accomplish: $L$ goes to $M_2$ and forwards a message to the rear particle. When the rear particle gets the message, it orders its predecessor to become the new rear particle and forward an ``all done'' message to $L$. From this point on, the old rear particle will stop following the trail, and no other particle will ever communicate with it again.
\item The deployment of a particle that will expand in direction $d$ is similar to the previous item. The only difference is that, after the rear particle has transferred its role to its predecessor, it also expands in direction $d$. The direction is encoded by taking the forward direction of the trail as $0$ and numbering all other directions from $1$ to $5$ in clockwise order. There can be no misunderstanding, since we are assuming that all particles have already reached an agreement on a common handedness.
\item When the termination operation is reached, the simulation is over, and a \emph{dismantling procedure} has to be executed, which will be described next.
\end{enumerate}

\subsection{Dismantling the machine}
There is one last problem to solve: when the pivot $P$ has been deployed, only three particles are left to deploy: $L$, $M_1$, and $M_2$. Since three particles are too few to simulate a tracing RAM, they have no way to reach their final locations without ``losing the way''. It is impossible to adapt our algorithm without making an extra assumption on the shape to be formed.

First we give two definitions. Recall that the algorithm $A''$ generates all the movement operations that are necessary to draw the principal subset $C'_f$ of the configuration $C_f$, by tracing it one scanline at a time. The last point plotted by $A''$ on the last scanline visited is said to be the \emph{last dot} of $C'_f$ (see Fig.~\ref{f:form02}). A subset of the grid graph $G$ is called a \emph{neighborhood} of point $p$ with radius $r$ if it contains precisely all the vertices of $G$ that have distance at most $r$ from $p$.

\begin{assumption}\label{a:main}
For each level $S_n$ of the shape, there exists a configuration $C_f$ of $n$ particles that forms $S_n$ (such that $C_f$ is unbreakably $k$-symmetric if $S_n$ has to be formed from an unbreakably $k$-symmetric initial configuration, cf.~Proposition~\ref{p:necessary}) and, for each principal subset $C'_f$, there exists a neighborhood of the last dot, with radius independent of $n$, that contains at least four particles of $C'_f$.
\end{assumption}

If we make Assumption~\ref{a:main} on the shape to be formed, we can complete our algorithm with a dismantling procedure, which makes $L$, $M_1$, $M_2$, and $P$ reach their final positions without getting lost in the attempt.

To this end, we must first modify the tracing algorithm $A''$ that we previously designed, and produce a new tracing algorithm $A'''$. Let $p_1$ be the particle occupying the last dot of $C'_f$, and let $p_2$, $p_3$, $p_4$ be the three particles of $C'_f$ closest to $p_1$ (other than $p_1$ itself). Ties are broken arbitrarily. The new algorithm $A'''$ proceeds exactly like $A''$, except that it skips the deployment operations (i.e., the operations of type~4 or~5) for $p_1$, $p_2$, $p_3$, $p_4$, and terminates when the ``pen'' reaches $p_1$.

Then we construct another algorithm $F$ that, with input $n$, generates the paths that three particles have to take from $p_1$ to reach the locations of $p_2$, $p_3$, $p_4$, making sure to avoid the locations of the other particles of $C_f$ (this is possible because we chose $p_2$, $p_3$, $p_4$ to be closest to $p_1$). Additionally, if some of the $p_i$s are expanded, $F$ outputs this information, as well as the direction of expansion. It is clear that $F$ is computable and terminates in a finite amount of time.

As before, we observe that there is a Turing machine $T$ that executes $A'''$, and a RAM $Q_T$ that simulates $T$. However, this time we add an extra register to $Q_T$, which will store the input value $n$ and will never erase it: this value will be passed as input to $F$. Again, the mobile mRAM $U_T$ is constructed, which is simulated by a 2-register RAM $Q'_{U_T}$, which in turn is simulated by particles $L$, $M_1$, $M_2$, and $P$.

When these four particles are done executing $A'''$, they erase all registers except the one containing $n$, and execute a simulation of $F$ (again, by simulating a 2-register RAM that executes $F$). Whenever $F$ outputs a step in a path from $p_1$ to $p_2$, $p_3$, or $p_4$, this step is memorized by the leader $L$. At the end of the simulation, $L$ has the three complete paths stored in its memory. Note that this is possible even if the memory of $L$ is of constant size: by Assumption~\ref{a:main}, the three paths are contained in a neighborhood of $p_1$ with radius independent of $n$, and therefore their length is bounded by a constant.

When the simulation of $F$ is terminated, $L$ starts moving toward $P$, and orders $M_1$ and $M_2$ to do the same. When $M_1$, $L$, $M_2$, and $P$ are in four consecutive locations, $L$ communicates the three paths to the others. At this point, each of the four particles knows what to do to reach its final position, and they all do so in an orderly fashion.

\subsection{Conclusion}

Our shape-formation algorithm demonstrates the following:

\begin{theorem}\label{t:main}
Under Assumption~\ref{a:main}, any computable shape is formable from any simply connected initial configuration.
\end{theorem}

The proof of correctness is straightforward, and it relies on two basic facts:
\begin{itemize}
\item A four-particle simulator and its trail of particles will never get in the way of other four-particle simulators. This is because each of them stays in its own principal sector; the only exceptions are the trespassers, which are actually not an obstacle due to the trespasser-avoidance rule.
\item The particles that have already been deployed will not get in the way of the four-particle simulator that deployed them. This is because the movement operations always make the simulator travel through locations where no particle has been deployed, yet. In particular, the delayed-deployment rule serves this purpose: a deployed particle will always expand toward a location that will never be traversed by the simulator again.
\end{itemize}

Note that Assumption~\ref{a:main} only excludes shapes whose levels are very sparse around the last dots of their principal subsets. In particular, connected shapes abundantly satisfy the assumption, which in this case reduces to the necessary condition of Proposition~\ref{p:necessary}. Therefore, we have the following characterization of formable connected computable shapes:

\begin{corollary}
A necessary and sufficient condition for a connected computable shape to be formable from a simply connected initial configuration is that, if the initial configuration is unbreakably $k$-symmetric, then also the corresponding level of the shape is unbreakably $k$-symmetric.
\end{corollary}
There are many possible ways to weaken Assumption~\ref{a:main}. For instance, we may require that the last four particles be in a constant-radius neighborhood of any point, not necessarily the last dot. Reaching such a point and computing the paths of the last four particles may be tricky, because other deployed particles may get in the way, but in principle it is possible. Actually, such extraneous particles may temporarily join the simulator, and allow it to compute final locations that are not necessarily contained in a neighborhood of constant radius.

Another way to remove Assumtpion~\ref{a:main} is to modify the requirements of the Shape-Formation problem, allowing a few particles to leave the system (or self-distruct) when the shape has been formed. This would neatly resolve the issue of how to dismantle the mRAM simulator upon termination, and it would only ``waste'' at most nine particles (three particles for each principal subset). Exploring these and other improvements to Theorem~\ref{t:main} is left as a direction for future research.

\section{Concluding remarks}

We have shown that four particles can simulate a mobile RAM, and we have applied such a construction to build a truly universal shape-formation algorithm. Our algorithm allows the formation of complex and general shapes that were not possible with previous approaches. 
Our investigations open several questions that are related to the specific Shape-Formation problem, as well as the general applicability of our mobile RAM. Concerning the Shape-Formation problem, it remains open to determine what happens when the starting configuration is not simply connected, and if our technical assumption for non-connected final shapes can be relaxed. As for our proposed mobile RAM, it would be interesting to understand what other applications can be found and what other problems can be effectively tackled with our tool.

\bibliographystyle{plainurl}
\bibliography{biblio}

\begin{thebibliography}{10}

\bibitem{ArCDRR18}
M.A. Arroyo, S.~Cannon, J.J. Daymude, D.~Randall, and A.W. Richa.
\newblock A stochastic approach to shortcut bridging in programmable matter.
\newblock {\em Natural Computing}, 17(5):343--365, 2018.

\bibitem{xCaDRR16}
S.~Cannon, J.J. Daymude, D.~Randall, and A.W. Richa.
\newblock A {Markov} chain algorithm for compression in self-organizing
  particle systems.
\newblock In {\em Proc.\ of the Symposium on Principles of Distributed
  Computing (PODC)}, pages 279--288, 2016.

\bibitem{xDaGPRSS18}
J.J. Daymude, Z.~Derakhshandeh, R.~Gmyr, A.~Porter, A.W. Richa, C.~Scheideler,
  and T.~Strothmann.
\newblock On the runtime of universal coating for programmable matter.
\newblock {\em Natural Computing}, 1:81--96, 2018.

\bibitem{xDaGPR+17}
J.J. Daymude, R.~Gmyr, A.W. Richa, C.~Scheideler, and T.~Strothmann.
\newblock Improved leader election for self-organizing programmable matter.
\newblock In {\em 13th International Symposium on Algorithms and Experiments
  for Wireless Sensor Networks, (ALGOSENSORS)}, pages 127--140, 2017.

\bibitem{DaHRS19}
J.J. Daymude, K.~Hinnenthal, A.W. Richa, and C.~Scheideler.
\newblock Computing by programmable particles.
\newblock In {\em {\em P. Flocchini, G. Prencipe, N. Santoro (Eds.):}
  Distributed Computing by Mobile Entities}. Springer, 2019.

\bibitem{xDerGMR+15}
Z.~Derakhshandeh, R.~Gmyr, A.W. Richa, C.~Scheideler, and T.~Strothmann.
\newblock An algorithmic framework for shape formation problems in
  self-organizing particle systems.
\newblock In {\em Proc.\ of NanoCom}, pages 21:1--21:2, 2015.

\bibitem{xspaa}
Z.~Derakhshandeh, R.~Gmyr, A.W. Richa, C.~Scheideler, and T.~Strothmann.
\newblock Universal shape formation for programmable matter.
\newblock In {\em Proc.\ of the 28th ACM Symposium on Parallelism in Algorithms
  and Architectures (SPAA)}, pages 289--299, 2016.

\bibitem{xDerGMR+17}
Z.~Derakhshandeh, R.~Gmyr, A.W. Richa, C.~Scheideler, and T.~Strothmann.
\newblock Universal coating for programmable matter.
\newblock {\em Theoretical Computer Science}, 671:56--68, 2017.

\bibitem{DerGSB+15}
Z.~Derakhshandeh, R.~Gmyr, T.~Strothmann, R.A. Bazzi, A.W. Richa, and
  C.~Scheideler.
\newblock Brief announcement: On the feasibility of leader election and shape
  formation with self-organizing programmable matter.
\newblock In {\em Proc.\ of ACM Symposium on Principles of Distributed
  Computing (PODC)}, pages 67--69, 2015.

\bibitem{xDerGSB+15}
Z.~Derakhshandeh, R.~Gmyr, T.~Strothmann, R.A. Bazzi, A.W. Richa, and
  C.~Scheideler.
\newblock Leader election and shape formation with self-organizing programmable
  matter.
\newblock In {\em Proc.\ of the International Conference on DNA Computing and
  Molecular Programming}, pages 117--132, 2015.

\bibitem{dilunaline}
G.~A. {Di Luna}, P.~Flocchini, N.~Santoro, and G.~Viglietta.
\newblock {TuringMobile: A Turing Machine of Oblivious Mobile Robots with
  Limited Visibility and Its Applications}.
\newblock In {\em Proc. of the International Symposium on Distributed Computing
  (DISC)}, pages 19:1--19:18, 2018.

\bibitem{previous}
G.A. Di~Luna, Paola Flocchini, Nicola Santoro, Giovanni Viglietta, and Yukiko
  Yamauchi.
\newblock Shape formation by programmable particles.
\newblock {\em Distributed Computing}, 2019.

\bibitem{xlinerec}
G.A.~Di Luna, P.~Flocchini, G.~Prencipe, N.~Santoro, and G.~Viglietta.
\newblock Line recovery by programmable particles.
\newblock In {\em Proc.\ of the International Conference on Distributed
  Computing and Networking}, 2018.

\bibitem{xmichail2}
O.~Michail, G.~Skretas, and P.G. Spirakis.
\newblock On the transformation capability of feasible mechanisms for
  programmable matter.
\newblock In {\em Proc.\ of the International Colloquium on Automata, Languages
  and Programming (ICALP)}, pages 136:1--136:15, 2017.

\bibitem{Minsky}
M.L. Minsky.
\newblock {\em Computation: Finite and Infinite Machines}.
\newblock Prentice-Hall, Inc., 1967.

\bibitem{xnewref}
A.~Naz, B.~Piranda, J.~Bourgeois, and S.C. Goldstein.
\newblock A distributed self-reconfiguration algorithm for cylindrical
  lattice-based modular robots.
\newblock In {\em Proc.\ of the IEEE International Symposium on Network
  Computing and Applications}, pages 254--263, 2016.

\bibitem{michail}
M.~Othon.
\newblock Terminating distributed construction of shapes and patterns in a fair
  solution of automata.
\newblock {\em Distributed Computuing}, 31(5):343--365, 2018.

\bibitem{xPa14}
M.J. Patitz.
\newblock An introduction to tile-based self-assembly and a survey of recent
  results.
\newblock {\em Natural Computing}, 13(2):195--224, 2014.

\bibitem{6696971}
J.~W. {Romanishin}, K.~{Gilpin}, and D.~{Rus}.
\newblock M-blocks: Momentum-driven, magnetic modular robots.
\newblock In {\em Proc.\ of the International Conference on Intelligent Robots
  and Systems}, pages 253--263, 2013.

\bibitem{xRo06}
P.W. Rothemund.
\newblock Folding {DNA} to create nanoscale shapes and patterns.
\newblock {\em Nature}, 440(7082):297--302, 2006.

\bibitem{6224638}
M.~{Rubenstein}, C.~{Ahler}, and R.~{Nagpal}.
\newblock Kilobot: A low cost scalable robot system for collective behaviors.
\newblock In {\em Proc.\ of the International Conference on Robotics and
  Automation}, pages 3293--3298, 2012.

\bibitem{xScW15}
N.~Schiefer and E.~Winfree.
\newblock Universal computation and optimal construction in the chemical
  reaction network-controlled tile assembly model.
\newblock In {\em Proc.\ of the International Conference on DNA Computing and
  Molecular Programming}, pages 34--54, 2015.

\bibitem{xToM91}
T.~Toffoli and N.~Margolus.
\newblock Programmable matter: concepts and realization.
\newblock {\em Physica D}, 47(1):263--272, 1991.

\bibitem{xWaWA04}
J.E. Walter, J.L. Welch, and N.M. Amato.
\newblock Distributed reconfiguration of metamorphic robot chains.
\newblock {\em Distributed Computing}, 17(2):171--189, 2004.

\end{thebibliography}

\end{document}